\documentclass[epsfig,12pt,onecolumn]{article}
\usepackage{amsfonts}
\usepackage{amssymb}
\usepackage{amsmath}
\usepackage{multicol}
\usepackage{graphicx}
\usepackage{float}
\usepackage{caption}

\makeatletter \@addtoreset{equation}{section}

\def \be{\begin{equation}}
\def \ee{\end{equation}}
\def \bea{\begin{eqnarray}}
\def \eea{\end{eqnarray}}

\newcommand{\nc}{\newcommand}
\nc{\al}{\alpha} \nc{\bib}{\bibitem} \nc{\la}{\lambda}
\nc{\C}{\mbox{\hspace{1.24mm}\rule{0.2mm}{2.5mm}\hspace{-2.7mm} C}}
\nc{\R}{\mbox{\hspace{.04mm}\rule{0.2mm}{2.8mm}\hspace{-1.5mm} R}}
\input{tcilatex}
\begin{document}

\title{\textbf{A proof of holographic complexity conjecture: wormhole
between horizon and singularity}}
\author{M. Bousder \\
{\small Faculty of Sciences, Mohammed V University in Rabat, Morocco}}
\maketitle

\begin{abstract}
This \textit{letter} provides evidence of complexity-volume and
complexity-action conjectures by examining the structure of a black hole,
which comprises a horizon linked to the singularity through a wormhole. In
this situation, the shape of the black hole's geometry resembles that of
Gabriel's horn. In essence, our results indicate that the information
paradox is merely the painter's paradox.

\textbf{Keywords: }Black Holes, Hawking radiation, information paradox.
\end{abstract}

\section{Introduction}

The Complexity-Volume (CV) \cite{HC1,HC5} conjecture proposes an association
between complexity \cite{HC0} and volume. According to this conjecture,
complexity can be understood as the volume of the maximum co-dimension one
surface that is attached to the boundary. We obtain a formula of the
complexity that reads CV-conjecture 
\begin{equation}
\mathcal{C}=\frac{\max V}{G_{N}L},  \label{m1}
\end{equation}%
where $V$ is the volume of the Einstein-Rosen bridge (ERB) and $L$ is the
radius of curvature of the bulk spacetime. The proposal known as
complexity-action (CA)-conjecture \cite{HC2} suggests that the quantum
computational complexity of a holographic state can be determined by
evaluating the on-shell action on a bulk region referred to as the Wheeler
De Witt (WDW) patch \cite{HC2} 
\begin{equation}
\mathcal{C}=\frac{\text{Action}}{\pi \hbar }.  \label{m2}
\end{equation}%
By interpreting the growth of complexity as computation and applying
normalization through above equation, we find that our proposals align with
the notion of Lloyd's conjectured bound for a system with energy $M$ \cite%
{NAT1}%
\begin{equation}
\frac{d\mathcal{C}}{dt}\leq \frac{2M}{\pi \hbar }.  \label{m3}
\end{equation}%
In the context of rotating black holes in 2+1-dimensional AdS, when the
ground state exhibits $M$ and angular momentum $J$ approaches zero, we have 
\cite{HC3}%
\begin{equation}
\frac{d\mathcal{C}}{dt}\leq \frac{2M}{\pi \hbar }\sqrt{1-\frac{J^{2}}{%
L^{2}M^{2}}},  \label{m4}
\end{equation}%
When considering electrically charged black holes in 3+1 dimensional AdS
that are significantly smaller than the AdS scale, we find that the minimum
of $M-\mu Q$, at a fixed $\mu $, occurs as $M$ and $Q$ approach zero \cite%
{HC2}%
\begin{equation}
\frac{d\mathcal{C}}{dt}\leq \frac{2M}{\pi \hbar }\sqrt{1-\frac{Q^{2}}{%
G_{N}M^{2}}}.  \label{m5}
\end{equation}

\section{Wormhole between horizon and singularity}

\begin{figure}[tbp]
\centering\includegraphics[width=13cm]{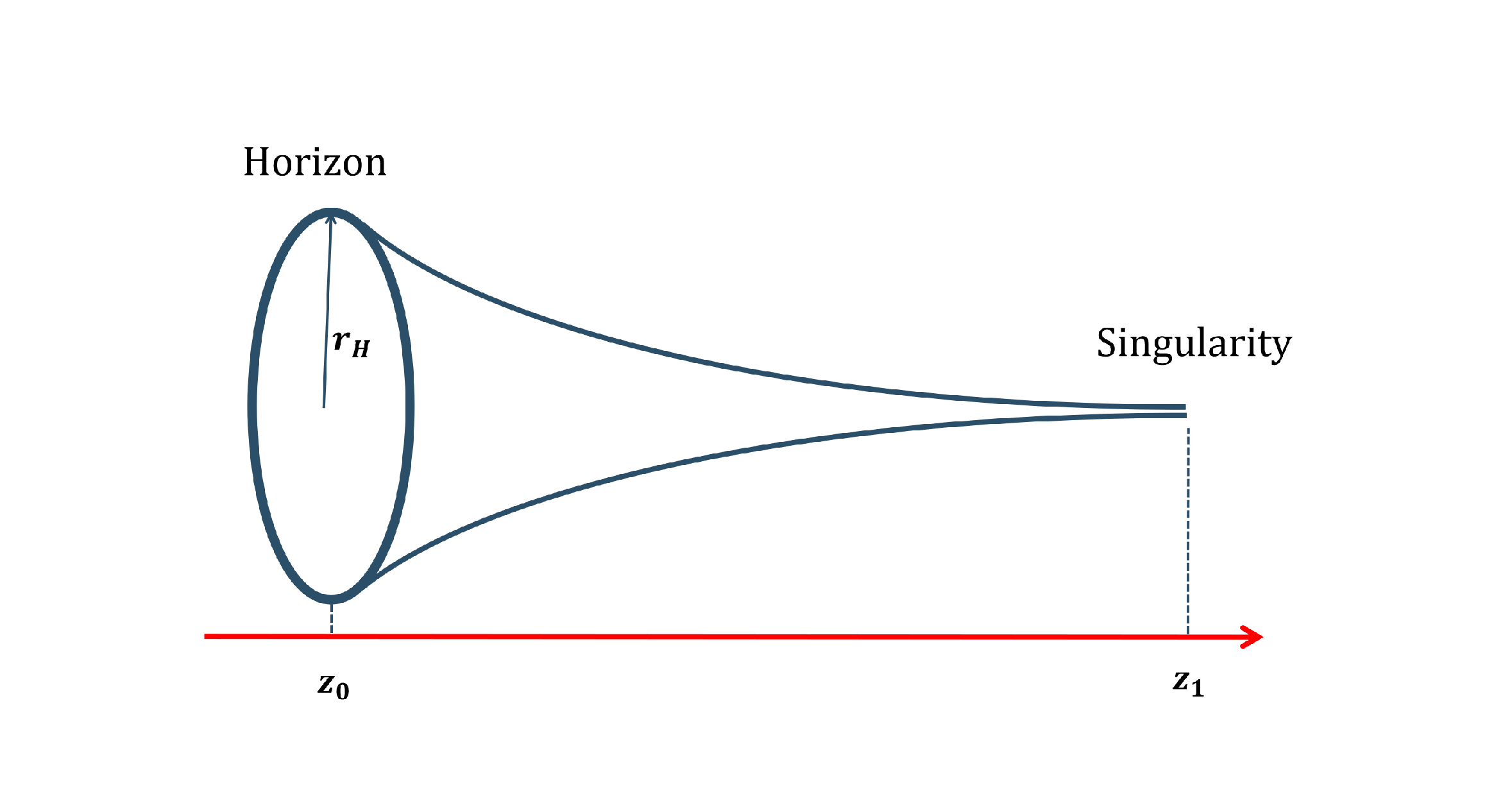}
\caption{In the case where $z_{0}-z_{1}$ approaches infinity, the appearance
of a black hole can take on a pseudosphere-like geometry, more precisely
resembling Gabriel's horn.}
\label{F1}
\end{figure}

We assume that a wormhole exists, connecting the singularity located at
point $z_{1}$ and the horizon at point $z_{0}$. In this scenario, if the
distance between the horizon and the singularity is very large, the
geometric structure of a black hole can be described as Gabriel's horn
shape, with the distance between points $z_{0}$ and $z_{1}$ representing the
length of this interconnected wormhole Figure (\ref{F1}). We consider only
two dimensions of the disk on the horizon. 

\subsection{Proof of the Complexity-Volume}

To create Gabriel's horn \cite{GH1,GH2}, one constructs the graph of $\frac{1%
}{z}$. By considering the domain $z\geq z_{0}$, the graph is rotated about
the $z$-axis in three dimensions to form pseudosphere. By employing
integration, one can determine both the volume $V$ and surface area $A$
using the concepts of solid of revolution and the surface of revolution. We
first introduce the volume of Gabriel's horn:%
\begin{equation}
V=\pi V_{0}L\int_{z_{0}}^{z_{1}}\left( \frac{1}{z^{2}}\right) ^{2}dz=\pi
V_{0}L\left( \frac{1}{z_{0}}-\frac{1}{z_{1}}\right) .  \label{a1}
\end{equation}%
Here $L$ is the radius of curvature of the $2+1$ dimensional bulk spacetime
and $V_{0}$\ is a parameter with a volume unit. The variable $z_{1}$ can
take on arbitrarily large values, as evident from the equation. However, it
is apparent that the volume of the segment of the horn bounded by $z_{0}$
and $z_{1}$ will never surpass $\pi $. Nevertheless, as the value of $z_{1}$
increases, the volume gradually converges towards $\pi $. This behavior can
be expressed using the limit of the singularity approaches infinity:%
\begin{equation}
\max V=\lim_{z_{1}\rightarrow \infty }V=\frac{\pi V_{0}L}{z_{0}}.  \label{a2}
\end{equation}%
From this expression, we have%
\begin{equation}
V_{0}=\frac{z_{0}}{\pi L}\max V.  \label{a3}
\end{equation}%
As the value of $V_{0}$ is not dependent on $z$ but rather on another
parameter, we select the value of $V_{0}$ accordingly: $V_{0}=r_{h}^{2}z_{%
\ast }$, where $\pi r_{h}^{2}$ represents the area of the surface of the
horizon disc at point $z_{0}$. Here, $z_{\ast }$ is a parameter that has the
unit of length, such that $\pi r_{h}^{2}z_{\ast }$ is the volume of a
cylinder. The cylinder's description adheres to the holographic principle,
highlighting the crucial parameter $z_{\ast }$. This parameter imparts the
black hole its holographic properties. From this and the area law $S=\frac{A%
}{4G_{N}}=\frac{\pi r_{h}^{2}}{G_{N}}$, we obtain for black holes in $2+1$
dimensional bulk%
\begin{equation}
\frac{z_{\ast }}{z_{0}}S=\frac{\max V}{G_{N}L}.  \label{a4}
\end{equation}%
$\max V=\frac{4\pi r_{H}^{3}}{3}$%
\begin{equation}
\frac{4\pi r_{H}^{3}}{3G_{N}L}=\frac{4r_{H}}{3L}S_{BH},
\end{equation}%
\begin{equation}
\frac{4r_{H}}{3L}S_{BH}=\mathcal{C},
\end{equation}%
\begin{eqnarray*}
z_{\ast } &=&4r_{H} \\
z_{0} &=&3L
\end{eqnarray*}%
This relationship is in good agreement with the CV-conjecture (\ref{m1}). In
addition, entropy is always linked to complexity \cite{GH3,GH4}, in this
case the term $\mathcal{C}=\frac{z_{\ast }}{z_{0}}S$ is only the complexity.
Currently, the nature of the $z$ parameter remains undetermined. However, in
the next subsequent section, we demonstrate that $z$ represents a time
parameter.

\subsection{Surface of pseudosphere}

The surface of Gabriel's horn is written as \cite{GH1,GH2}%
\begin{equation}
A=2\pi A_{0}\int_{z_{0}}^{z_{1}}\frac{1}{z}\sqrt{1+\left( \frac{z_{0}}{z}%
\right) ^{4}}dz.  \label{b1}
\end{equation}%
where $A_{0}$ is a parameter to the dimension of the surface. The following
condition is given: 
\begin{equation}
A_{\ast }=2\pi A_{0}\int_{z_{0}}^{z_{1}}\frac{dz}{z}=2\pi A_{0}\ln \frac{%
z_{1}}{z_{0}}\leq A  \label{b2}
\end{equation}%
We can determine the limit of $A$ (\ref{b1}) from $A_{\ast }$ such that the
singularity approaches infinity: $\lim_{z_{1}\rightarrow \infty }A\geq
\lim_{z_{1}\rightarrow \infty }A_{\ast }=\infty $. Therefore,%
\begin{equation}
\lim_{z_{1}\rightarrow \infty }A=\infty .  \label{b3}
\end{equation}%
Based on this connection, an infinite number of horizons exist in the $z$%
-direction. This relationship illustrates that the surface of the wormhole,
extending from the horizon to the singularity, can store an infinite amount
of information than on volume (\ref{a2}), which is in good agreement with
the holographic principle \cite{HC6}. The apparent paradox lies in the fact
that Gabriel's horn has a finite volume but an infinite surface area. This
seems contradictory because we might expect that a solid with an infinite
surface area would also have an infinite volume. However, the resolution
lies in understanding the nature of infinity. From (\ref{a2}) and (\ref{b3})
we have%
\begin{equation}
\frac{\pi z_{\ast }}{z_{0}}S=\frac{\max V}{G_{N}L}\text{,\ vs \ }%
\lim_{z_{1}\rightarrow \infty }A=\infty .  \label{b4}
\end{equation}%
These two relationships indicate the equivalence between the information
paradox and the painter's paradox. Using (\ref{b1}) we find%
\begin{equation}
\frac{dA}{dz}=2\pi \frac{A_{0}}{z}\sqrt{1+\left( \frac{z_{0}}{z}\right) ^{4}}%
.  \label{b5}
\end{equation}%
We recall that $V_{0}=r_{h}^{2}z_{\ast }$, for this we offer that $%
A_{0}=2r_{h}z_{\ast }$ and using the \ the Schwarzschild radius $%
r_{h}=2G_{N}M$ for $c=1$, we find%
\begin{equation}
\frac{1}{4\pi G_{N}}\frac{dA}{dz_{\ast }\ln z}=2M\sqrt{1+\left( \frac{z_{0}}{%
z}\right) ^{4}}.  \label{b6}
\end{equation}%
We recall that $z_{\ast }$ is constant which depends on the length of the
Gabriel's horn. The complexity must maintain a continuous growth pattern
determined by \cite{HC4} $\frac{d\mathcal{C}_{A}}{dt}\sim ST$, where $T$ is
the temperature. If we compare this equation with (\ref{m4}) and (\ref{m5}),
we notice that $z_{\ast }\ln z$ plays the role of time, for this we assume $%
t\equiv z_{\ast }\ln z$. From (\ref{a4}) we obtain $\frac{d\mathcal{C}}{dt}=%
\frac{d\max V}{G_{N}Ldt}\sim X\frac{dA}{dt}$, with $X$ is a parameter with
the unit of length. This leads to%
\begin{equation}
\frac{d\mathcal{C}}{dt}\sim \frac{2M}{\pi \hbar }\sqrt{1+\left( \frac{z_{0}}{%
z}\right) ^{4}}.  \label{b7}
\end{equation}%
This expression is equivalent to (\ref{m3}), (\ref{m4}) and (\ref{m5}).

\section{Conclusion}

In conclusion, this \textit{letter} presents compelling evidence supporting
the complexity-volume and complexity-action conjectures by analyzing the
intricate structure of a black hole. By investigating the connection between
the horizon and singularity through a wormhole, we find that the geometry of
a black hole closely resembles that of Gabriel's horn. These intriguing
findings shed light on the nature of black holes and offer insights into the
information paradox. Moreover, our research suggests that the information
paradox, a long-standing puzzle in the field, may be analogous to the
painter's paradox. Just as a painter can create an infinite amount of
information on a finite volume of the horn, the information within a black
hole might be preserved despite its seemingly limited volume. This notion
challenges traditional notions of information conservation and opens up new
avenues for further exploration.

\end{document}